\documentclass[a4paper]{mem}
\usepackage{natbib}
\usepackage{graphicx}
\begin{document}
   \title{Evolution of AGN - an optical view
}

   \author{C. Wolf }

   \institute{Department of Physics, Oxford University,
               Keble Road, Oxford, OX1 3RH, U.K.   }

   \abstract{
   This article discusses the our changing knowledge of the evolution of
   AGN from an optical perspective. It focusses on optically-unobscured
   (type-1) AGN as it is much trickier to investigate the evolution type-2
   objects from an optical point of view. It discusses major optical survey
   work that improved our knowledge of QSO evolution during the last five
   years. It touches on the shape of the QSO luminosity function, on the
   peak of QSO activity, the quest for reddened QSOs and spectral evolution.
   Finally, it summarizes recent advances in the research of host galaxies,
   which probably hold the key for the triggering mechanisms of the activity,
   and reiterates the difficulty of understanding the physical evolution of
   AGN and its place in the context of galaxy evolution.
   }
   \authorrunning{C. Wolf}
   \titlerunning{Evolution of AGN - optical view}
   \maketitle

\section{Introduction}

Shortly after the discovery of quasars, it was understood that these objects are truely remarkable, not only in their physical nature but also in their cosmological evolution. Schmidt (1968, 1970) observed a very sharp decline in the space density of quasars, over roughly a factor of 100 from $z\sim 2$ to $z\sim 0$. This decline was so remarkable, that it could be credibly demonstrated from a sample as small as 20 objects. The interpretation was supported by $V/V_{\rm max}$ tests, developed for this purpose at the time. The strong evolution clearly applied to both, objects selected optically and those selected from their radio emission. At the time it suggested that quasars were the most dramatically evolving population of objects in the Universe.

It was later understood that quasars are only one part of a larger population of objects commonly called Active Galactic Nuclei or AGN. From an optical perspective, a defining characteristic of AGN is non-stellar continuum radiation. More sensitive observations showed that also light from luminous and young stars is sometimes found in AGN, suggesting that star formation may be occuring in the host galaxy and may have a common trigger with the nuclear activity. AGN have been observed over a wide range of luminosities which presently covers eight orders of magnitude, across $M_B\approx [-10,-30]$. This range poses a challenge for observations aiming at understanding the phenomenon: at the lowest luminosities the host galaxy outshines the active nucleus, making it hard to detect. At the highest luminosities, the AGN outshines the host galaxy, making the latter almost impossible to detect, let alone to investigate. Traditionally, unobscured AGN at $M_B<-23$ have been called QSOs. This definition does not reflect any physical limits, but is motivated by observational practicalities: here the AGN is luminous enough to dominate the combined spectrum over the host galaxy, making its AGN nature easy to recognize.

This article focusses on the more luminous half, $M_B<-20$, of the range covered by AGN, where $>95$\% of the total luminosity density is emitted. Besides these bright unobscured {\it type-1} objects, there is an entire class of optically obscured {\it type-2} AGN with a very significant contribution to the cosmic black hole accretion history. However, being optically obscured makes these objects hard to find and study from an optical point of view, although their host galaxies are clearly observable. Hence, this article will give more weight to type-1 objects, here called QSOs irrespective of their luminosity.

\section{QSO Evolution -- 5 Years Ago}

A useful instrument to quantify the activity statistics of AGN is their luminosity function (LF). Only five years ago, there was still confusion about the evolutionary pattern of the type-1 LF. While a strong increase of the activity out to $z\sim 2$ had been established a long time before, and a turnover must be trivially expected at higher redshift somewhere, the shape of the turnover and the redshift of any peak in activity was still unclear. Boyle, Shanks \& Peterson (1988) described the LF at $z=[0.3,2.2]$ with a broken power-law, characterized by the parameters $L^*(z)$ and $\phi^*(z)$. These allowed to distinguish between simple evolutionary models such as Pure Luminosity Evolution (PLE) and Pure Density Evolution (PDE), with PLE being the preferred description for the low-redshift ($z<2$) evolution of QSOs. At high redshift $z=[2.7,4.7]$ Schmidt, Schneider \& Gunn (1995) parametrized the LF with a single power-law that does not feature any break luminosity, because their small sample and narrow luminosity range did not constrain any more parameters. In their $L$-range they found a strong decrease in the activity when going towards higher redshift. Finally, Warren, Hewitt \& Osmer (1994) bridged the gap and investigated the turnover within $z=[2.0,4.5]$. Their model fit followed a complicated description of luminosity-dependent density evolution (LDDE) with a strong high-redshift cutoff at $z>3.3$.

While the rise and fall of quasars between the Big Bang and today was established by these observations, there was still disagreement on the shape and redshift of the turnover. E.g., at $z=[2.2,3.6]$ the selection of QSOs from colours was difficult due to abundant stars having similar colours as the rare QSOs, and hence quantifying the selection function carefully was nearly impossible. Furthermore, each QSO sample at $z>2$ contained only a few dozen objects across a narrow luminosity range, which left the shape of the LF unconstrained (for details see Osmer 2003 and references therein). Clearly, larger surveys mapping the luminosity-redshift plane as widely as possible were needed to advance the field of QSO evolution.

\section{Recent Optical Surveys}

A number of surveys were carried out to overcome all the shortcomings of previous optical surveys for QSOs, such as difficult selection functions, narrow luminosity ranges and small samples. I list six important and complementary surveys here, which have produced results in recent years:
\begin{enumerate}
\item the brightest ($\sim 400$) objects were observed in the wide-area objective-prism Hamburg-ESO-Survey (HES, Wisotzki 2000) at $B<17$, $z=[0.0,3.2]$ 
\item the 2dF Quasar Redshift Survey (2QZ, Boyle et al. 2000, Croom et al. 2004) found large numbers ($\sim 23,000$) of QSOs at $z=[0.3,2.2]$ and $B<21$.
\item the Sloan Digital Sky Survey (SDSS, York et al. 2000) collects the largest QSO sample to date with important complete subsamples at $z>3.6$, $i<20$ (Fan et al. 2001)
\item the COMBO-17 survey (Wolf et al., 2003) selected QSOs reliably at $z=[1,5]$ with detailed SEDs from 17 filters, providing the deepest large sample
\item the BTC-40 survey (Monier et al. 2001) targetted specifically the very high-z end at $z>4.8$, confirming just two of these rare QSOs to date
\item a Lyman-break-selected galaxy sample (Steidel et al. 2002) providing a small but the deepest probe into nuclear activity at $z\sim 3$ (Hunt et al. 2004)
\end{enumerate}

All these surveys selected their object samples by colour, and all except for HES and COMBO-17 assembled a QSO sample by spectroscopic follow-up of a subset. In contrast, HES and COMBO-17 had their spectroscopic information already obtained in the first observational step: HES is based on object prism spectroscopy and collapsed spectra into colours for selection; the COMBO-17 filters have sufficient number and spectral resolution to identify QSOs and measure accurate redshifts from photometry alone.

Indeed, the enlarged statistics on colours and spectra of QSOs allowed to develop the technique of photometric redshifts for QSOs to maturity: broad-band colours such as the SDSS $ugriz$ bands lead to photo-z errors of $\sigma_z \sim 0.2$ or $\sim 0.1$ in photo-z codes based on QSO templates (Richards et al. 2001) or using neural nets (Budavari et al. 2001), respectively. In the COMBO-17 filter set the larger number of filters leads to a much more complete identification of QSOs, especially in the range of the suspected turnover at $z=[2,4]$. The higher spectral resolution of medium-band filters leads to much smaller photo-z errors of $\sigma_z / (1+z) \approx 0.015$ for QSOs (Wolf et al. 2004). Similar work in the CADIS survey did not take variability into account, and the resulting variation in observed colours led to a number of redshift outliers there (Wolf et al. 1999). The latter comparison demonstrates the importance of simultaneous colours for photometric redshifts of QSOs.

The analysis of these surveys has resulted in LFs covering different parts of the redshift-luminosity plane, and we have found that space densities mostly agree well between surveys where redshift and luminosity ranges overlap. Especially, it was shown that the COMBO-17 LF is consistent with 2QZ at low redshift and smoothly turns over into the SDSS LF at high redshift (see Fig.~1 and Wolf et al. 2003), providing the previously missing link.

\subsection{Shape of the luminosity function}

Where luminosity functions from different surveys overlap, they tend to agree. However, the parts of the redshift-luminosity plane measured by more than one survey are not large, and there are still some remaining issues about survey incompleteness and its correction. 

The largest newly observed sample is from 2QZ and covers redshifts of $z=[0.3,2.2]$. The resulting LFs can be fit to broken power laws and give satisfactory fits for a PLE-type model. However, 2QZ selects candidates from broad-band colours and thus has a  bias against low-luminosity QSOs with bright host galaxies, where the latter dominate the colours. Some of the flattening of the towards the faint end may result from this incompleteness, and it is not clear that a broken power law is the required function.

At the brightest end, the HES has found some mild curvature in the LF. At very low redshifts of $z<0.3$, it should clearly allow to see the $L^*$ knee in the LF suggested by the early broken power-laws and seen again by the 2QZ survey. However, while there is curvature in the LF and flattening towards lower luminosities, a broken power-law does not match the HES results, which instead suggest a steeper faint LF than 2QZ.

At high redshifts of $z>3.6$ the SDSS sample has improved over the original SSG95 result. The results do not allow to constrain LF curvature due to small sample size and a narrow luminosity range, yet. However, the slope of the SDSS high-z LF is flatter than the 2QZ slope at lower redshift, suggesting evolution in the LF shape.

The COMBO-17 sample bridged the gap between 2QZ and SDSS with complete selection, and furthermore extended the LFs to much fainter limits. The COMBO-17 LF shows again mild curvature but no $L^*$ break in its low-luminosity range. Combined with the SDSS LF, COMBO-17 shows a flattening of the high-z LF, as a faint extrapolation of the single power-law SDSS LF exceeds the COMBO-17 counts.

The Steidel et al. Lyman-break sample provides the deepest probe, although the object number does not help to constrain the faint-end slope or LF curvature reliably. It was argued, that this sample combined with the WHO94 results suggested an $L^*$ break with a strong slope change. However, combining the larger samples from COMBO-17 and SDSS gives consistent results in terms of a smoothly curved LF without a distinct break.

Altogther, these results suggest that the LF of QSOs is smoothly curved, but we have not seen a compelling case for a strong $L^*$ break, which could be traced as a clue to the physical evolution of QSOs underlying our observations.

   \begin{figure}
   \centering
   \resizebox{\hsize}{!}{\rotatebox[]{-90}{\includegraphics[clip=true]{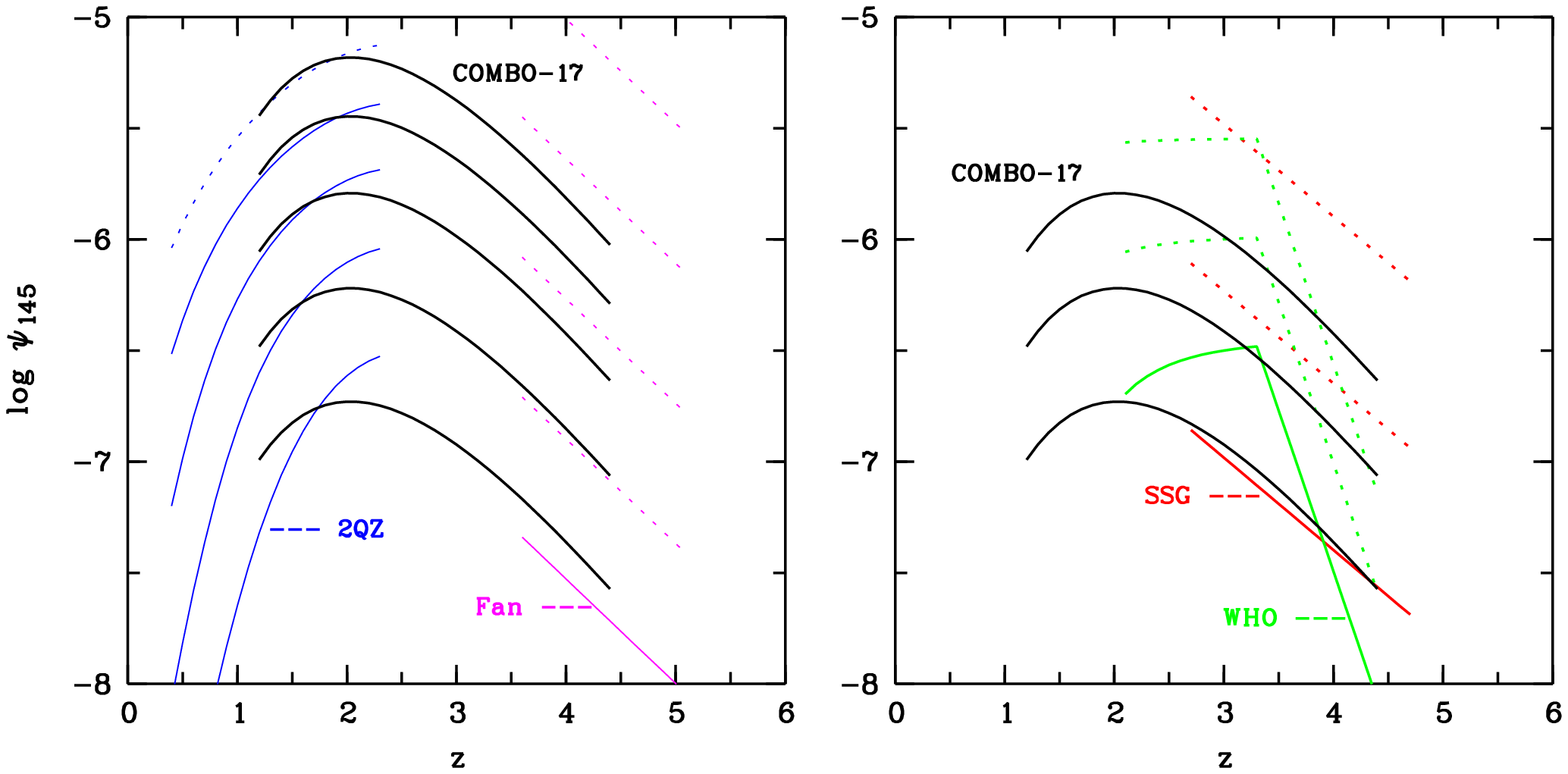}}}
   \caption{QSO space densities from 2QZ, COMBO-17 and SDSS (Fan et al. 2001) agree
	very well. The different curves show densities according to the best-fitting 
	models from each survey, integrated to different limiting luminosities of
	$M_{\rm 145,Vega} < [-28\ldots-24]$. Dashed lines are extrapolations
	to fainter magnitudes.}
              \label{QSOspacedens}%
    \end{figure}

\subsection{Peak of AGN activity}

The peak of the cosmic AGN activity would be measured best by the integrated AGN luminosity density, which could potentially be used as a crude proxy for the total BH accretion, once the conversion becomes clearer. The faint COMBO-17 observations have demonstrated that the $L$-range of existing samples probably covers $>95$\% of the total blue AGN luminosity density. Thus, variations in the further faint extrapolation would change the total luminosity density only little. COMBO-17 found an activity peak at $z\approx 2.0$ based on its LF at $M_B<-22$. In contrast, the much shallower HES found the luminosity density to rise further out to its survey limit at $z=3.2$ from an LF integrated at $M_B<-28$. This comparison clarifies that high-luminosity QSOs reached their activity peak earlier than low-luminosity objects and implies that the shape of the QSO LF changes with time. The same behaviour was also found in X-ray selected samples, where low-$L_x$ AGN peak at $z\sim 0.75$ (Cowie et al. 2003, Steffen et al. 2003). This phenomenon is called {\it cosmic downsizing} and describes a scenario, in which comparatively large members of a population undergo the critical evolutionary epoch first, with members of a progressively smaller kind showing up only later. This downsizing trend provides an important constraint for theories of co-evolution between AGN and their host galaxies.

\subsection{Reddened QSOs}

Aside from the subject of optically-obscured QSOs, the optical community occasionally wondered about missing red or reddened QSOs which may escape colour selection or show unexpected spectral shapes. Recently, a careful study of a large SDSS QSO sample in combination with 2MASS photometry investigated the detailed continuum properties of QSOs at $z<2.2$ (Richards et al. 2003, Hopkins et al. 2004). This work showed that the QSO population could best be explained by a distribution of intrinsic spectral slopes $\alpha=[-0.75,-0.25]$, combined with dust reddening at the redshift of the QSO, possibly with an SMC-type dust law. It was shown that the AGN continuum and the broad-line region (BLR) were reddened to the same degree but the narrow-line region (NLR) remained unaffected. The mean reddening was measured to be $\langle E_{B-V} \rangle = 0.03$, while the SDSS selection procedure was shown to be sensitive to $E_{B-V}<0.5$. Only 2\% of the parent population of type-1 QSOs was found to be reddened with $E_{B-V}>0.1$. A total of $\sim10$\% of QSOs are lost from the SDSS sample by extinction pushing objects below the flux limit, not due to reddening-induced colour changes!

\subsection{Optically obscured QSOs}

The canonical model of AGN involves a dusty torus around the accreting black hole, which absorbs optical signatures of the central engine on certain lines-of-sight, while keeping a type-1 appearance at other viewing angles. Among low-luminosity AGN, this picture had been confirmed through the unification of Seyfert-1 galaxies with Seyfert-2 galaxies. The key observation in the latter was a detection of unobscured AGN light, which was scattered (and polarized) by low-density dust residing far above the absorbing torus plane and being freely illuminated by the AGN. Seyfert-2 galaxies had been found abundantly as counterparts to Seyfert-1 galaxies, but only with low luminosity. Five years ago, optically obscured QSOs (with high luminosity) were still at large, although a radio galaxy at $z=0.44$, allegedly the most powerful IRAS source, was a good candidate for a type-2 QSO (Kleinmann et al. 1988) and further X-ray-based evidence was obtained by Franceschini et al. (2000).

More recently, very significant progress was made with deep X-ray surveys and optical identification of faint X-ray sources by spectroscopic follow-up. Several type-2 QSOs were found with Chandra and XMM, reaching up to $z=3.7$ (Norman et al. 2002). It has been confirmed that type-2 objects dominate the population at low luminosity, but their fraction drops when going to high luminosities. At this point, it seems unclear which population dominates the nuclear accretion budget, because of the debate on how to correct for completeness in follow-up identification. E.g., Treister et al. (this volume) suggest, that with proper correction type-2 objects account for 3/4 of the integrated X-ray luminosity even at $z>2$, where we only see higher-$L_x$ objects.

Maybe somewhat unexpectedly, the subject of optically obscured AGN has been advanced through extensive work based on optical selection in the SDSS spectroscopic database. Kauffmann et al. (2003) have selected type-2 AGN at $z<0.3$ by subtracting the stellar continuum and star formation-related components from emission lines to isolate the AGN emission from the NLR. Zakamska et al. (2003, 2004) have pushed further in redshift by applying such a technique to the whole SDSS spectral database. They isolated a serendipitous sample at $z=[0.3,0.8]$ with $\sim 150$ type-2 QSOs, defined by $L_{\rm O {\sc III}}>3\times 10^8 L_{\rm sol}$, and many lower-luminosity objects. Their AGN nature was further confirmed by showing that this class of objects is MIR/FIR-luminous and has hard X-ray colours. However, in contrast to the Kauffmann et al. sample at $z<0.3$, the statistical completeness of the Zakamska et al. sample is presently unclear.

While X-ray surveys played a fundamental role in mapping out the obscured population, they still need to improve in area, and optical/NIR follow-up needs to reach deeper, before the obscured samples at high-z will catch up in usefulness with type-1 samples. We will need a good $z,L$-map of the type-1/2 fraction, not only to observe the entire nuclear accretion budget, but also in order to constrain models of torus evolution.

\subsection{Evolution of spectral properties}

Any evolution in spectral properties of QSOs would hold important clues about changes in the physical conditions in AGN, which would help very much to interpret the AGN phenomenon beyond the above discussed 'bean counting' exercises of the many surveys. However, no significant redshift evolution has been observed in the properties of either the continuum or the emission lines. Pentericci et al. (2003) complemented the optical data of 45 SDSS QSOs at $z=[3.6,5.0]$ with JHK photometry and found a mean spectral slope of $\alpha = -0.57\pm 0.33$, which is fully consistent with the QSOs at low redshift. Emission lines were found to display various (anti-) correlations between their equivalent widths (EWs) and the continuum luminosity of the source, also known as Baldwin effects. But no evolution was found for EWs with redshift, leaving us with no evidence for chemical evolution (Croom et al. 2002, Dietrich et al. 2002). Also, no evolution with redshift has been found in a relation between black hole mass and QSO luminosity, $M_{\rm BH} \propto L_{\rm QSO}$ (see Corbett et al. 2003 and references therein).

\section{Host galaxies of AGN}

The field of AGN host galaxies has seen great progress over the last few years, and is the most promising approach to understand which phenomenon triggers nuclear activity. Type-1 QSOs have typically been studied with HST, because any bright seeing-enlarged central light source swamps the light of the host galaxy, making the delicate subtraction of the overpowering AGN source impossible. Adaptive optics may become an alternative ground-based technique, but suffer currently still from significant and variable wings in their PSF. Type-2 AGN appear like pure host galaxies in the optical and NIR and are hence trivial to examine.

Kukula et al. (2001) and Dunlop et al. (2003) targetted type-1 QSOs of luminosity $M_V<-23.5$ with NICMOS at $z=[0.9,1.9]$ and with WFPC2 at $z<0.2$, respectively. These monochromatic images showed host galaxies to be normal giant ellipticals obeying the normal Kormendy relation expected at their redshifts.

Kauffmann et al. (2003) examined type-2 AGN at $z<0.3$ in the SDSS spectroscopic sample and found normal-looking early-type (spheroidal) galaxies as well as some disks and disturbed systems. At lowest AGN luminosities these exhibit the usual red colours expected in early-type galaxies, but at higher luminosity they show a mild excess of blue star light compared to non-AGN early-type objects. Detailed diagnostics using the $H\delta$ absorption line suggest that most of them experienced a starburst within the previous Gyr. 

Most recently, two studies on low-luminosity type-1 AGN ($M_V=[-24,-20]$) measured not only shapes but also colours of type-1 hosts. Jahnke et al. investigated a sample at $z<0.2$ and Sanchez et al. used a $z=[0.5,1.1]$ sample from the GEMS (Galaxy Evolution from Morphologies and SEDs) survey. They both found mostly spheroidal morphologies but also some disks. Again, many early-types showed a moderate excess of blue star light compared to non-AGN.

These studies appear to show consistently, that AGN live mostly (but not exclusively) in 'young bulges', i.e. mostly large ellipticals with a moderate excess of blue stellar light compared to non-AGN ellipticals. This suggests that the AGN phenomenon is accompanied by some amount of star formation, which is either still ongoing or has predated the AGN phase as a recent starburst. This picture is not surprising, given that a powerful AGN requires both a massive black hole and abundant fuel supply to operate. While only massive early-type galaxies would contain such black holes, only galaxies with significant recent or ongoing star formation provide the fuel supply.

\section{AGN evolution in the context of galaxy evolution}

The hierarchical paradigm holds major mergers responsible for triggering starbursts and for the transport of gas to nuclear regions leading to AGN activity, with AGN feedback potentially heating and clearing away remaining gas from the host and terminating star formation. Incidentally, major mergers also lead to the formation of spheroidal galaxies, which will need a Gyr or more to settle onto the red sequence of non-star-forming, passively evolving galaxies. This picture is broadly consistent with the observations of AGN host galaxies.

Observations of a relationship between quiescent black-hole masses and the velocity dispersion of the hosting galaxy bulge (Ferrarese \& Merritt 2000; Gebhardt et al. 2000) have suggested a co-evolution of galaxies and AGN. This relation can be further developed into a relation between $M_{\rm BH}$ and stellar mass of the bulge. However, recently, Treu, Malkan \& Blandford (2004) found the $M_{\rm BH}$-$\sigma$ relation at $z=0.37$ offset from the one at $z=0$. If this result is confirmed, it implies that AGN were active and black holes have grown before the stellar bulges reached their final mass. On a currently more speculative side, high-resolution VLA observation made by Walter et al. (2004) support this view, claiming that the host galaxy of the highest-redshift QSO known, SDSS J1148+5251 at $z=6.42$, consists mostly of molecular gas, with very little room left for stellar mass, effectively ruling out the presence of a $10^{12}~M_{\rm sol}$ stellar bulge required by the local $M_{\rm BH}$-$\sigma$ relation.

\section{Summary}

In the recent five years observers have made great progress in mapping the cosmic nuclear accretion history. The luminosity function of type-1 QSOs has been mapped over ever larger ranges of the redshift-luminosity plane. We have reached reasonable consistency between different surveys and have constrained most of the UV-optical AGN luminosity density within the range of existing surveys. We have found no evidence of spectral evolution - AGN appear to be always the same kind of animal, irrespective of the epoch of observation.

The census of optically-obscured (type-2) AGN is starting to catch up, but there is still much work needed to understand their contribution at redshift $z>1$. Whether we look at obscured or unobscured AGN, at X-ray or optical luminosities, we probably see a pattern of cosmic downsizing among AGN, where high-luminosity AGN were the first active objects in the early Universe, while lower-luminosity AGN lagged behind and only became abundant later.

At redshift $z<2$ we see AGN of either type being hosted predominantly by elliptical galaxies with little morphological disturbance and some extra light from younger stars. They do not appear particularly disturbed and do not undergo simultaneous starbursts. In principle, AGN could be a late phase of major mergers where the nuclear activity only commences after the merger has relaxed into a regular elliptical, although this is not the only permitted interpretation.

On the whole, it is still unclear how to constrain the physical evolution of AGN and how to place them into the context of galaxy evolution. New estimates of black hole masses in the more distant Universe and observations of host galaxies at very high redshift might still harbor quite a few surprises for us.

\begin{acknowledgements}
	I am grateful to P. Osmer for comments improving the manuscript.
	This work was supported by a PPARC Advanced Fellowship.
\end{acknowledgements}

\bibliographystyle{aa}

\begin{thebibliography}{}

\bibitem[]{}
 Boyle, B.J., Shanks, T. \& Peterson, B.A. 1988, MNRAS, 235, 935
\bibitem[]{}
 Boyle, B.J., Shanks, T. et al. 2000, MNRAS, 317, 1014
\bibitem[]{}
 Budavari, T., Csabai, I., Szalay, A.S. et al. 2001, AJ, 122, 1163
\bibitem[]{}
 Corbett, E.A., Croom, S.M. et al. 2003, MNRAS, 343, 705
\bibitem[]{}
 Cowie, L.L. et al. 2003, ApJ, 584, L57
\bibitem[]{}
 Croom, S.M., Rhook, K. et al. 2002, MNRAS, 337, 275
\bibitem[]{}
 Croom, S.M., Smith, R.J. et al. 2004, MNRAS, 349, 1397
\bibitem[]{}
 Dietrich, M., Hamann, F. et al. 2002, ApJ, 581, 912
\bibitem[]{}
 Dunlop, J.S., McLure, R.J., Kukula, M.J. et al. 2003, MNRAS, 340, 1095
\bibitem[]{}
 Fan, X., Strauss, M.A., Schneider, D.P. et al. 2001, AJ, 121, 54
\bibitem[Ferrarese \& Merritt (2000)]{FM00} 
 Ferrarese, L. \& Merritt, D. 2000, ApJ, 539, L9
\bibitem[]{}
 Franceschini, A. et al. 2000, A\&A, 353, 910
\bibitem[Gebhardt et al. (2000)]{Geb00}
 Gebhardt, K. et al. 2000 ApJ, 539, L13 
\bibitem[]{}
 Hopkins, P.F., Strauss, M.A. et al. 2004, AJ, 128, 1112
\bibitem[Hunt et al. (2004)]{Hunt04}
 Hunt, M.P., Steidel, C.C. et al. 2004, ApJ, 605, 625
\bibitem[]{}
 Jahnke, K., Kuhlbrodt, B. \& Wisotzki, L. 2004, MNRAS, 352, 399
\bibitem[]{}
 Kauffmann, G., Heckman, T.M., Tremonti, C. et al. 2003, MNRAS, 346, 1055
\bibitem[]{}
 Kleinmann, S.G. et al. 1988, ApJ, 328, 161
\bibitem[]{}
 Kukula, M.J., Dunlop, J.S., McLure, R.J. et al. 2001, MNRAS, 326, 1533
\bibitem[]{}
 Monier, E.M., Kennefick, J.D. et al. 2002, AJ, 124, 2971
\bibitem[]{}
 Norman, C., Hasinger, G. et al. 2002, ApJ, 571, 218
\bibitem[]{}
 Osmer, P. 2003, in {\it Coevolution of black holes and galaxies}, L.C. Ho (ed.).
 Carnegie Observatories Astrophysics Ser., Vol. 1.
\bibitem[]{}
 Pentericci, L., Rix, H.-W., Prada, F. et al. 2003, A\&A, 410, 75
\bibitem[]{}
 Richards, G.T., Weinstein, M.A. et al. 2001, AJ, 122, 1151
\bibitem[]{}
 Richards, G.T., Hall, P.B., VandenBerk, D.E. et al. 2003, AJ, 126, 1131
\bibitem[]{}
 Sanchez, S.F., Jahnke, K., Wisotzki, L. et al. 2004, ApJ, 614, 586
\bibitem[]{}
 Schmidt, M. 1968, ApJ, 151, 393
\bibitem[]{}
 Schmidt, M. 1970, ApJ, 162, 371
\bibitem[]{}
 Schmidt, M., Schneider, D.P. \& Gunn, J.E. 1995, AJ, 110, 68
\bibitem[]{}
 Steffen, A.T. et al. 2003, ApJ, 596, L23
\bibitem[]{}
 Steidel, C.C., Hunt, M.P., Shapley, A.E. et al. 2002, ApJ, 576, 653
\bibitem[]{}
 Treister, E., Urry, M. et al. 2004, this volume
\bibitem[Treu et al. (2004)]{treu04}
 Treu, T., Malkan, M.A. \& Blandford, R.D. 2004, ApJ, 615, L97
\bibitem[Walter et al. (2004)]{Wal04}
 Walter, F. et al., 2004, ApJL, 615, L17
\bibitem[]{}
 Warren, S.J., Hewett, P.C. \& Osmer, P.S. 1994, ApJ, 421, 412
\bibitem[]{}
 Wisotzki, L. 2000, A\&A, 353, 853
\bibitem[]{}
 Wisotzki, L., Christlieb, N. et al. 2000, A\&A, 358, 77
\bibitem[]{}
 Wolf, C., Meisenheimer, K., R\"oser, H.-J. et al. 1999, A\&A, 343, 399
\bibitem[]{}
 Wolf, C., Wisotzki, L. et al. 2003, A\&A, 408, 499
\bibitem[]{}
 Wolf, C., Meisenheimer, K., Kleinheinrich, M. et al. 2004, A\&A, 421, 913
\bibitem[]{}
 York, D.G. et al. 2000, AJ, 120, 1579
\bibitem[]{}
 Zakamska, N.L., Strauss, M.A. et al. 2003, AJ, 126, 2125
\bibitem[]{}
 Zakamska, N.L., Strauss, M.A. et al. 2004, AJ, 128, 1002

\end{thebibliography}

\end{document}